\documentclass[superscriptaddress,nobibnotes,aps,prd,nofootinbib]{revtex4}%
\usepackage{amsmath}
\usepackage{amsfonts}
\usepackage{amssymb}
\usepackage{graphicx}%
\begin{document}
\title{Casimir Wormholes}
\author{Remo Garattini}
\email{remo.garattini@unibg.it}
\affiliation{Universit\`{a} degli Studi di Bergamo, Dipartimento di Ingegneria e Scienze
Applicate,Viale Marconi 5, 24044 Dalmine (Bergamo) Italy and }
\affiliation{I.N.F.N. - sezione di Milano, Milan, Italy.}

\begin{abstract}
Casimir energy is always indicated as a potential source to generate a
Traversable Wormhole. It is also used to proof the existence of negative
energy which can be built in the laboratory. However, in the scientific
literature there is no trace of the consequences on the traversable wormhole
itself. In this work, we would like to consider such a source to see if and
which kind of traversable wormhole can be produced. As a further analysis, we
examine also the consequences of Quantum Weak Energy Conditions on the
traversability of the wormhole. We find that an agreement with the original
Casimir traversable wormhole is found. Nevertheless, despite of the
traversability result, one finds once again that the traversability is only in
principle but not in practice. 

\end{abstract}
\maketitle

\section{Introduction}

The Casimir effect appears between two plane parallel, closely spaced,
uncharged, metallic plates in vacuum. It was predicted theoretically in
1948\cite{Casimir} and experimentally confirmed in the Philips
laboratories\cite{Sparnaay,AVS}. However, only in recent years further
reliable experimental investigations have confirmed such a
phenomenon\cite{Lamoreaux}. The interesting feature of this effect is that an
attractive force appears which is generated by negative energy. Indeed the
attractive force arises because the renormalized energy assumes the following
form%
\begin{equation}
E^{\text{Ren}}\left(  a\right)  =-\frac{\hbar c\pi^{2}S}{720a^{3}},
\end{equation}
where $S$ is the surface of the plates and $a$ is the separation between them.
The force can be obtained with the computation of%
\begin{equation}
F\left(  a\right)  =-\frac{dE^{\text{Ren}}\left(  a\right)  }{da}%
=-3\frac{\hbar c\pi^{2}S}{720a^{4}},
\end{equation}
producing also a pressure%
\begin{equation}
P\left(  a\right)  =\frac{F\left(  a\right)  }{S}=-3\frac{\hbar c\pi^{2}%
}{720a^{4}}.\label{P(a)}%
\end{equation}
It is immediate to recognize that the energy density is nothing but%
\begin{equation}
\rho_{C}\left(  a\right)  =-\frac{\hbar c\pi^{2}}{720a^{4}}\label{rhoC}%
\end{equation}
suggesting the existence of a relation between the pressure $P$ and the energy
density $\rho$ described by an Equation of State (EoS) of the form
$P=\omega\rho$ with $\omega=3$. The nature of this effect is connected with
the Zero Point Energy (ZPE) of the Quantum Electrodynamics Vacuum distorted by
the plates. It is important to observe that this effect has a strong
dependence on the geometry of the boundaries. Indeed, Boyer\cite{Boyer}
proofed the positivity of the Casimir effect for a conducting spherical shell
of radius $r$. The same positivity has been proofed also in Ref.\cite{BVW}, by
means of heat kernel and zeta regularization techniques. As far as we know,
the Casimir energy represents the only artificial source of \textit{exotic
matter}\ realizable in a laboratory\footnote{Actually, there exists also the
possibility of taking under consideration a squeezed vacuum. See for example
Ref.\cite{Hochberg}.}\textit{.} Exotic matter violates the Null Energy
Condition (NEC), namely for any null vector $k^{\mu}$, we have $T_{\mu\nu
}k^{\mu}k^{\nu}\geq0$. Violation of the NEC is related to the existence of a
bizarre but amazing object predicted by General Relativity: a
\textit{Traversable Wormhole}. Traversable wormholes (TW) are solutions of the
Einstein's Field Equations (EFE) powered by classical sources\cite{MT,Visser}.
However, given the quantum nature of the Casimir effect, the EFE must be
replaced with the semiclassical EFE, namely%
\begin{equation}
G_{\mu\nu}=\kappa\left\langle T_{\mu\nu}\right\rangle ^{\text{Ren}}%
\qquad\kappa=\frac{8\pi G}{c^{4}},
\end{equation}
where $\left\langle T_{\mu\nu}\right\rangle ^{\text{Ren}}$ describes the
renormalized quantum contribution of some matter fields: in this specific
case, the electromagnetic field. To establish a connection between
\textit{exotic matter} and a TW, we introduce the following spacetime metric%
\begin{equation}
ds^{2}=-e^{2\phi(r)}\,dt^{2}+\frac{dr^{2}}{1-b(r)/r}+r^{2}\,(d\theta^{2}%
+\sin^{2}{\theta}\,d\varphi^{2})\,,\label{metric}%
\end{equation}
representing a spherically symmetric and static wormhole. $\phi(r)$ and $b(r)$
are arbitrary functions of the radial coordinate $r\in\left[  r_{0}%
,+\infty\right)  $, denoted as the redshift function, and the shape function,
respectively \cite{MT,Visser}. A fundamental property of a wormhole is that a
flaring out condition of the throat, given by $(b-b^{\prime}r)/b^{2}>0$, must
be satisfied as well as the request that $1-b(r)/r>0$. Furthermore, at the
throat $b(r_{0})=r_{0}$ and the condition $b^{\prime}(r_{0})<1$ is imposed to
have wormhole solutions. It is also fundamental that there are no horizons
present, which are identified as the surfaces with $e^{2\phi}\rightarrow0$, so
that $\phi(r)$ must be finite everywhere. With the help of the line element
$\left(  \ref{metric}\right)  $, we can write the EFE in an orthonormal
reference frame, leading to the following set of equations%
\begin{equation}
\frac{b^{\prime}\left(  r\right)  }{r^{2}}=\kappa\rho\left(  r\right)
,\label{rhoEFE}%
\end{equation}%
\begin{equation}
\frac{2}{r}\left(  1-\frac{b\left(  r\right)  }{r}\right)  \phi^{\prime
}\left(  r\right)  -\frac{b\left(  r\right)  }{r^{3}}=\kappa p_{r}\left(
r\right)  ,\label{pr0}%
\end{equation}%
\begin{align}
&  \Bigg\{\left(  1-\frac{b\left(  r\right)  }{r}\right)  \left[  \phi
^{\prime\prime}\left(  r\right)  +\phi^{\prime}\left(  r\right)  \left(
\phi^{\prime}\left(  r\right)  +\frac{1}{r}\right)  \right]  \nonumber\\
&  -\frac{b^{\prime}\left(  r\right)  r-b\left(  r\right)  }{2r^{2}}\left(
\phi^{\prime}\left(  r\right)  +\frac{1}{r}\right)  \Bigg\}=\kappa
p_{t}(r),\label{pt0}%
\end{align}
in which $\rho\left(  r\right)  $ is the energy density\footnote{However, if
$\rho\left(  r\right)  $ represents the mass density, then we have to replace
$\rho\left(  r\right)  $ with $\rho\left(  r\right)  c^{2}.$}, $p_{r}\left(
r\right)  $ is the radial pressure, and $p_{t}\left(  r\right)  $ is the
lateral pressure. We can complete the EFE with the expression of the
conservation of the stress-energy tensor which can be written in the same
orthonormal reference frame%
\begin{equation}
p_{r}^{\prime}\left(  r\right)  =\frac{2}{r}\left(  p_{t}\left(  r\right)
-p_{r}\left(  r\right)  \right)  -\left(  \rho\left(  r\right)  +p_{r}\left(
r\right)  \right)  \phi^{\prime}\left(  r\right)  .\label{Tmn}%
\end{equation}
The purpose of this paper is to establish if the Casimir energy can be really
considered as a source for a TW. In Ref.\cite{MTY}, this proposal was examined
concluding that the plate separation should be smaller than the electron
Compton wavelength that it means that the physical apparatus cannot be
realized. Later Visser in his book\cite{Visser} proposed a realistic model for
the total Stress-Energy Tensor (SET) represented by%
\begin{equation}
T_{\sigma}^{\mu\nu}=\sigma\hat{t}^{\mu}\hat{t}^{\nu}\left[  \delta\left(
z\right)  +\delta\left(  z-a\right)  \right]  +\Theta\left(  z\right)
\Theta\left(  a-z\right)  \frac{\hbar c\pi^{2}}{720a^{4}}\left[  \eta^{\mu\nu
}-4\hat{z}^{\mu}\hat{z}^{\nu}\right]  ,\label{TCas}%
\end{equation}
where $\hat{t}^{\mu}$ is a unit time-like vector, $\hat{z}^{\mu}$ is a normal
vector to the plates and $\sigma$ is the mass density of the plates. He
concluded that the mass of the plates compensates so much the negative energy
density forbidding therefore the creation of a TW. Nonetheless, in both
Refs.\cite{MTY, Visser} and also \cite{FR}, nothing has been said to the
possible form of the shape function $b\left(  r\right)  $ and the related
redshift function $\phi\left(  r\right)  $. Since an EoS is implicitly
working, namely $p_{r}\left(  r\right)  =\omega\rho\left(  r\right)  $, with
$\omega=3$, one is tempted to impose the simplest choice on the redshift
function, namely $\phi^{\prime}\left(  r\right)  =0$. This assumption yields
the following form for the metric%
\begin{equation}
ds^{2}=-\,dt^{2}+\frac{dr^{2}}{1-\left(  \frac{r_{0}}{r}\right)  ^{\frac{4}%
{3}}}+r^{2}\,(d\theta^{2}+\sin^{2}{\theta}\,d\varphi^{2})\,\label{EoS}%
\end{equation}
leading to the following SET%
\begin{equation}
T_{\mu\nu}=\frac{1}{\kappa r^{2}}\left(  \frac{r_{0}}{r}\right)
^{\frac{\omega+1}{\omega}}diag\left(  -\frac{1}{\omega},-1,\frac{\omega
+1}{\omega},\frac{\omega+1}{\omega}\right)
\end{equation}
which, for $\omega=3$, has not any connection with the original Casimir SET.
Thus it becomes important to explore other possibilities which can lead to the
best correspondence between a TW and the Casimir SET. For instance, one can
examine the consequences of the constraint imposed by a \textit{Quantum Weak
Energy Condition} (QWEC)%
\begin{equation}
\rho(r)+p_{r}(r)=-f\left(  r\right)  \qquad f\left(  r\right)  >0\qquad\forall
r\in\left[  r_{0},+\infty\right)  .\label{EoSb}%
\end{equation}
The QWEC has been introduced for the first time in Ref.\cite{PMMMV} and
subsequently used in cosmology in Ref.\cite{MBLAEPMMTOYV}.\bigskip\ As far as
we know, in the context of traversable wormholes the first appearance of the
QWEC $\left(  \ref{EoSb}\right)  $ is in Ref.\cite{BLLMM}, with%
\begin{equation}
f\left(  r\right)  =A\left(  r_{0}/r\right)  ^{\alpha},\label{f(r)}%
\end{equation}
where $A$ is an appropriate constant introduced to describe an energy density.
Note that the condition $\left(  \ref{EoSb}\right)  $ has a direct connection
with the \textit{volume integral quantifier}, which provides information about
the \textit{total amount}\ of averaged null energy condition (ANEC) violating
matter in the spacetime\cite{VKD}. This is defined by%
\begin{equation}
I_{V}=\int[\rho(r)+p_{r}(r)]dV
\end{equation}
and for the line element $\left(  \ref{metric}\right)  $, one can write%
\begin{equation}
I_{V}=\frac{1}{\kappa}\int\left(  r-b\left(  r\right)  \right)  \left[
\ln\left(  \frac{e^{2\phi(r)}}{1-\frac{b\left(  r\right)  }{r}}\right)
\right]  ^{\prime}dr,\label{IV}%
\end{equation}
where the measure $dV$ has been changed into $r^{2}dr$. For example, the
calculation of $I_{V}$ for the traversable wormhole\cite{MT}%
\begin{equation}
ds^{2}=-dt^{2}+\frac{dr^{2}}{1-r_{0}/r}+r^{2}\,(d\theta^{2}+\sin^{2}{\theta
}\,d\varphi^{2})\,,
\end{equation}
leads to%
\begin{equation}
I_{V}=-\frac{1}{\kappa}\int\left(  r-r_{0}\right)  \left[  \ln\left(
1-\frac{r_{0}}{r}\right)  \right]  ^{\prime}dr=-\frac{r_{0}}{\kappa}\left[
\ln\left(  r\right)  \right]  _{r_{0}}^{+\infty}\rightarrow-\infty
.\label{IVmw0}%
\end{equation}
This expression is diverging when $r\rightarrow\infty$ indicating that an
infinite amount of exotic matter is required to maintain the wormhole. We may
conclude that this result prohibits the existence of \textit{zero tides}\ and
\textit{zero density}\ wormholes. The interesting feature of the QWEC $\left(
\ref{EoSb}\right)  $ is that $b\left(  r\right)  $ can be determined exactly,
not only for the form introduced in Ref.\cite{BLLMM}, but even for a generic
$f\left(  r\right)  $. Indeed, with the help of Eqs.$\left(  \ref{rhoEFE}%
\right)  $ and $\left(  \ref{pr0}\right)  $, we can write%
\begin{equation}
\frac{b^{\prime}\left(  r\right)  }{r}+\left[  2\left(  1-\frac{b\left(
r\right)  }{r}\right)  \phi^{\prime}\left(  r\right)  -\frac{b\left(
r\right)  }{r^{2}}\right]  =-\kappa rf\left(  r\right)  \label{p+rho}%
\end{equation}
and, following Ref.\cite{BLLMM}, we introduce the following auxiliary function
$u\left(  r\right)  =1-b\left(  r\right)  /r$ leading to%
\begin{equation}
u^{\prime}\left(  r\right)  -2u\left(  r\right)  \phi^{\prime}\left(
r\right)  =-\kappa rf\left(  r\right)  .
\end{equation}
This equation can be easily integrated and the form of the solution is%
\begin{equation}
b\left(  r\right)  =r\left[  1-\kappa\exp\left(  2\phi\left(  r\right)
\right)  \int_{r_{0}}^{r}\exp\left(  -2\phi\left(  r^{\prime}\right)  \right)
f\left(  r^{\prime}\right)  r^{\prime}dr^{\prime}\right]  ,\label{bprho}%
\end{equation}
where the integration constant was fixed taking into account the throat
condition. Therefore fixing the form of $\phi\left(  r\right)  $, one can
determine $b\left(  r\right)  $. As an example, we briefly report the constant
case examined in Ref.\cite{BLLMM}. We find%
\begin{gather}
b\left(  r\right)  =r\left[  1-\kappa\exp\left(  2\phi\left(  r\right)
\right)  A\int_{r_{0}}^{r}\exp\left(  -2\phi\left(  r^{\prime}\right)
\right)  r^{\prime}dr^{\prime}\right]  \nonumber\\
=r\left(  1-\frac{\kappa A}{2}\left(  r^{2}-r_{0}^{2}\right)  \right)  ,
\end{gather}
where we have set $f\left(  r\right)  =A$, and we have also considered the
additional \textit{Zero Tidal Force }(ZTF) condition. It is immediate to
recognize that the previous shape function does not represent a traversable
wormhole\footnote{See Ref.\cite{BLLMM} for further details.}. It is also
possible to apply the reverse procedure, namely we fix the form of the shape
function and we compute the redshift function and what we obtain is%
\begin{equation}
\phi\left(  r\right)  =\phi\left(  r_{0}\right)  +\frac{1}{2}\int_{r_{0}}%
^{r}\left[  \frac{b\left(  \bar{r}\right)  }{\bar{r}^{2}}-\frac{b^{\prime
}\left(  \bar{r}\right)  }{\bar{r}}-\kappa\bar{r}f\left(  \bar{r}\right)
\right]  \left(  1-\frac{b\left(  \bar{r}\right)  }{\bar{r}}\right)
^{-1}d\bar{r}.\label{phiprho}%
\end{equation}
For instance, in the simple case of a constant $f\left(  r\right)  $ and a
constant $b\left(  r\right)  $, we find%
\begin{equation}
\phi\left(  r\right)  =\phi\left(  r_{0}\right)  +\frac{1}{2}\int_{r_{0}}%
^{r}\left[  \frac{r_{0}}{\bar{r}^{2}}-\kappa\bar{r}f\right]  \left(
1-\frac{r_{0}}{\bar{r}}\right)  ^{-1}d\bar{r}\underset{r\rightarrow r_{0}%
}{\simeq}\ln\left(  r-r_{0}\right)
\end{equation}
and also in this case, we have not a TW, rather a black hole. However, we will
see that in some cases, the QWEC offers interesting forms for the shape
function and for the redshift function, respectively. The rest of the paper is
structured as follows, in section \ref{p2} we continue the investigation to
determine if the Casimir energy density $\left(  \ref{rhoC}\right)  $ can be
considered as a source for a traversable wormhole, in section \ref{p3} we use
the QWEC to determine which kind of shape function we can obtain, in section
\ref{p4} we determine under which condition the QWEC will produce a
traversable wormhole. We summarize and conclude in section \ref{p5}. Units in
which $\hbar=c=k=1$ are used throughout the paper.

\section{The Casimir Traversable Wormhole}

\label{p2}In this section we assume that our exotic matter will be represented
by the Casimir energy density $\left(  \ref{rhoC}\right)  $. Following
Ref.\cite{MTY}, we promote the constant plates separation $a$ to a radial
coordinate $r$. Even if the authors of Ref.\cite{MTY} assume a Casimir device
made by spherical plates, we have to say that the SET form they use is the
same obtained with the flat plates assumption. Therefore the curvature of the
plates introduces some modification which, in this first approximation, will
be neglected. We have also to observe that the replacement of $a$ with $r$
could make the stress-energy tensor $\left(  \ref{TCas}\right)  $, with
$\sigma=0$, potentially not conserved as expected. Our strategy begins with
the examination of Eq.$\left(  \ref{rhoEFE}\right)  $ leading to the following
form of the shape function%
\begin{equation}
b\left(  r\right)  =r_{0}-\frac{\pi^{3}}{90}\left(  \frac{\hbar G}{c^{3}%
}\right)  \int_{r_{0}}^{r}\frac{dr^{\prime}}{r^{\prime2}}=r_{0}+\frac{\pi
^{3}l_{p}^{2}}{r90}-\frac{\pi^{3}l_{p}^{2}}{r_{0}90}=r_{0}-\frac{r_{1}^{2}%
}{r_{0}}+\frac{r_{1}^{2}}{r};\qquad r_{1}^{2}=\frac{\pi^{3}l_{p}^{2}}%
{90}\label{b(r)0}%
\end{equation}
where the throat condition $b(r_{0})=r_{0}$ has been imposed. To see if the
Casimir energy really generates a TW, we need to study the redshift function
which obeys Eq.$\left(  \ref{pr0}\right)  $. Plugging the shape function
$\left(  \ref{b(r)0}\right)  $ into the radial pressure $\left(
\ref{pr0}\right)  $, we find%
\begin{equation}
\frac{2}{r}\left(  1-\frac{r_{0}}{r}+\frac{r_{1}^{2}}{r_{0}r}-\frac{r_{1}^{2}%
}{r^{2}}\right)  \phi^{\prime}\left(  r\right)  -\frac{r_{0}}{r^{3}}%
+\frac{r_{1}^{2}}{r_{0}r^{3}}-\frac{\left(  1-\omega\right)  r_{1}^{2}}{r^{4}%
}=0,
\end{equation}
where we have used the EoS $p_{r}\left(  r\right)  =\omega\rho\left(
r\right)  $ and Eq.$\left(  \ref{rhoEFE}\right)  $. The solution for a generic
$\omega$ is
\begin{equation}
\phi\left(  r\right)  =-\frac{1}{2}\left[  \left(  \omega r_{0}^{2}-r_{1}%
^{2}\right)  {\frac{\ln\left(  r_{0}r+r_{1}^{2}\right)  }{r_{0}^{2}+r_{1}^{2}%
}}+\left(  1-\omega\right)  \ln\left(  r\right)  +\left(  \omega r_{1}%
^{2}-r_{0}^{2}\right)  {\frac{\ln\left(  r-r_{0}\right)  \omega\,r_{1}^{2}%
}{r_{0}^{2}+r_{1}^{2}}}\right]  +\phi\left(  r_{0}\right)  .\label{Phi(r)}%
\end{equation}
Since we have some freedom in fixing $\omega$ and $r_{0}$, we observe that
when $\omega r_{1}^{2}-r_{0}^{2}>0$, we find a black hole, while when we
fix\footnote{When $\omega r_{1}^{2}-r_{0}^{2}<0$, we obtain a singularity.}%
\begin{equation}
\omega r_{1}^{2}-r_{0}^{2}=0\qquad\Longrightarrow\qquad\omega=\omega_{0}%
=\frac{r_{0}^{2}}{r_{1}^{2}},\label{or1r0}%
\end{equation}
we obtain a TW. Assuming the last choice, we can write%
\begin{equation}
\phi\left(  r\right)  =\frac{1}{2}\left(  {\omega-1}\right)  {\ln\left(
\frac{r\left(  \omega+1\right)  }{\left(  \omega r+r_{0}\right)  }\right)
+\phi\left(  r_{0}\right)  ,}\label{Phi(r)1}%
\end{equation}
and for $r\rightarrow\infty$,%
\begin{equation}
\phi\left(  r\right)  \rightarrow\frac{1}{2}\left(  {\omega-1}\right)
{\ln\left(  \frac{\omega+1}{\omega}\right)  +\phi\left(  r_{0}\right)  ,}%
\end{equation}
therefore, in order to obtain the appropriate asymptotic flat limit, it is
convenient to fix\footnote{By imposing ${\phi\left(  r_{0}\right)  =0}$, the
redshift function should be%
\begin{equation}
\phi\left(  r\right)  =\frac{1}{2}\left(  {\omega-1}\right)  {\ln\left(
\frac{r\left(  \omega+1\right)  }{\left(  \omega r+r_{0}\right)  }\right)  .}%
\end{equation}
}%
\begin{equation}
{\phi\left(  r_{0}\right)  =-}\frac{1}{2}\left(  {\omega-1}\right)
{\ln\left(  \frac{\omega+1}{\omega}\right)  .}%
\end{equation}
Then, the redshift function reduces to%
\begin{equation}
\phi\left(  r\right)  =\frac{1}{2}\left(  {\omega-1}\right)  {\ln}\left(
{\frac{r\omega}{\left(  \omega r+r_{0}\right)  }}\right)  \label{Phi(r)2}%
\end{equation}
and the shape function becomes%
\begin{equation}
b\left(  r\right)  =\left(  1-\frac{1}{\omega}\right)  r_{0}+\frac{r_{0}^{2}%
}{\omega r}.\label{b(r)1}%
\end{equation}
The last component of the SET we have to compute is $p_{t}$. With the help of
Eq.$\left(  \ref{pt0}\right)  $, we find%
\begin{equation}
p_{t}(r)=\omega_{t}\left(  r\right)  \left(  \frac{r_{0}^{2}}{\kappa
\omega\,{r}^{4}}\right)  =\omega_{t}\left(  r\right)  \rho(r){{,}}%
\end{equation}
where we have introduced a inhomogeneous EoS on the transverse pressure with%
\begin{equation}
\omega_{t}\left(  r\right)  =-\frac{{\omega}^{2}\left(  4r-r_{0}\right)
+r_{0}\left(  4\omega+1\right)  }{4\left(  \omega\,r+r_{0}\right)
},\label{otr}%
\end{equation}
and the final form of the SET is%
\begin{equation}
T_{\mu\nu}=\frac{r_{0}^{2}}{\kappa\omega\,{r}^{4}}\left[  diag\left(
-1,-\omega,1,1\right)  +\left(  \omega_{t}\left(  r\right)  -1\right)
diag\left(  0,0,1,1\right)  \right]  .\label{Tmn1}%
\end{equation}
The conservation of the SET is satisfied but a comparison with the Casimir SET
shows that there is an extra contribution on the transverse pressure: this is
the consequence of having a function of the radial coordinate in front of the
SET instead of a constant. It is interesting to observe that the first tensor
is the Casimir SET iff $\omega=3$. Moreover when%
\begin{equation}
\frac{r_{0}^{2}}{\kappa\omega}=\frac{r_{1}^{2}}{\kappa}=\frac{c^{4}}{8\pi
G}\frac{\pi^{3}}{90}\left(  \frac{\hbar G}{c^{3}}\right)  =\frac{\hbar
c\pi^{2}}{720},
\end{equation}
the identification is complete. Combining the redshift function $\left(
\ref{Phi(r)2}\right)  $ with the shape function $\left(  \ref{b(r)1}\right)
$, we can write the line element $\left(  \ref{metric}\right)  $ in the
following way%
\begin{equation}
ds^{2}=-{\left(  \frac{r\omega}{\left(  \omega r+r_{0}\right)  }\right)
}^{{\omega-1}}\,dt^{2}+\frac{dr^{2}}{1-\frac{1}{r}\left(  \left(  1-\frac
{1}{\omega}\right)  r_{0}+\frac{r_{0}^{2}}{\omega r}\right)  }+r^{2}%
\,(d\theta^{2}+\sin^{2}{\theta}\,d\varphi^{2})\,\label{dSC2}%
\end{equation}
and for the special value $\omega=3$, we find%
\begin{equation}
\phi\left(  r\right)  =\ln\left(  \frac{3r}{3r+r_{0}}\right)  \qquad
and\qquad{b}\left(  r\right)  =\frac{2r_{0}}{3}+\frac{r_{0}^{2}}%
{3r},\label{phi(r)b(r)}%
\end{equation}
whose line element is
\begin{equation}
ds^{2}=-{\left(  \frac{3r}{3r+r_{0}}\right)  }^{2}\,dt^{2}+\frac{dr^{2}%
}{1-\frac{2r_{0}}{3r}-\frac{r_{0}^{2}}{3r^{2}}}+r^{2}\,(d\theta^{2}+\sin
^{2}{\theta}\,d\varphi^{2})\,,\label{dSC3}%
\end{equation}
describing a TW of Planckian size\footnote{$r_{0}=\sqrt{3}r_{1}\simeq
1.\,\allowbreak016\,6l_{P}$}. The corresponding SET $\left(  \ref{Tmn1}%
\right)  $ becomes%
\begin{equation}
T_{\mu\nu}=\frac{r_{0}^{2}}{3\kappa{r}^{4}}\left[  diag\left(
-1,-3,1,1\right)  +\left(  \frac{6r}{3r+r_{0}}\right)  diag\left(
0,0,1,1\right)  \right]  .\label{Tmn2}%
\end{equation}
Note that the inhomogeneous function $\left(  \ref{otr}\right)  $ is such that%
\begin{equation}
r\in\left[  r_{0},+\infty\right)  \qquad\Longrightarrow\qquad\omega_{t}\left(
r\right)  \in\left[  -\frac{5}{2},-3\right)
\end{equation}
and the transverse pressure on the throat becomes%
\begin{equation}
p_{t}\left(  r_{0}\right)  =\frac{5}{6\kappa{r}_{0}^{2}}.
\end{equation}
It is interesting also to observe that if $\omega=1$, we obtain the
Ellis-Bronnikov (EB) wormhole of sub-planckian size\footnote{$r_{0}%
=\sqrt{\frac{\pi^{3}}{90}}l_{p}=\allowbreak0.586\,95l_{p}$}%
\cite{ellisGL,Bronnikov}. Even in this case, the SET is conserved, but to
establish a connection with the Casimir SET, we need to write the SET of
$\left(  \ref{Tmn1}\right)  $ in the following way%
\begin{equation}
T_{\mu\nu}=\frac{r_{0}^{2}}{\kappa{r}^{4}}\left[  diag\left(
-1,-1,1,1\right)  \right]  =\frac{r_{0}^{2}}{\kappa{r}^{4}}\left[  diag\left(
-1,-3,1,1\right)  +2diag\left(  0,1,0,0\right)  \right]  ,\label{TEB1}%
\end{equation}
or in an equivalent representation\footnote{The SET is represented in an
orthonormal frame.}%
\begin{equation}
T_{\mu\nu}=\frac{r_{0}^{2}}{\kappa{r}^{4}}\left[  diag\left(
-1,-1,1,1\right)  \right]  =\frac{r_{0}^{2}}{2\kappa{r}^{4}}\left[
diag\left(  -1,-3,1,1\right)  +diag\left(  -1,1,1,1\right)  \right]
.\label{TEB2}%
\end{equation}
Both representations $\left(  \ref{TEB1}\right)  $ and $\left(  \ref{TEB2}%
\right)  $ show the traceless Casimir SET, but only the decomposition $\left(
\ref{TEB2}\right)  $ is obtained in a canonical way. In the next section, we
will explore the consequences of assuming a QWEC to see its relationship with
the Casimir energy and its connection with the formation of a TW.

\section{Global Monopoles from QWEC}

\label{p3}As pointed out in \cite{PMMMV}, the Casimir energy and the QWEC have
a strong connection. Therefore, it appears interesting which kind of shape
function we can extract by imposing such a profile on the relationship leading
to Eq.$\left(  \ref{bprho}\right)  $. By assuming a vanishing redshift
function, we can write%
\begin{equation}
b\left(  r\right)  =r\left[  1-\kappa\int_{r_{0}}^{r}f\left(  r^{\prime
}\right)  r^{\prime}dr^{\prime}\right]  \label{b(r)QWEC}%
\end{equation}
and with the help of $\left(  \ref{f(r)}\right)  $ one gets%
\begin{equation}
b\left(  r\right)  =r\left[  1-\kappa Ar_{0}^{\alpha}\int_{r_{0}}^{r}%
\frac{dr^{\prime}}{r^{\prime\alpha-1}}\right]  =r\left[  \left(
1-\frac{\kappa Ar_{0}^{\alpha}}{\alpha-2}\frac{1}{r_{0}^{\alpha-2}}\right)
+\frac{\kappa Ar_{0}^{\alpha}}{\alpha-2}\frac{1}{r^{\alpha-2}}\right]
,\qquad\alpha\neq2,\label{b(r)CGM}%
\end{equation}
while for $\alpha=2$, one finds%
\begin{equation}
b\left(  r\right)  =r\left[  1-\kappa Ar_{0}^{2}\int_{r_{0}}^{r}%
\frac{dr^{\prime}}{r^{\prime}}\right]  =r\left[  1-\kappa Ar_{0}^{2}\ln\left(
\frac{r}{r_{0}}\right)  \right]  .\label{b(r)CGMa}%
\end{equation}
The shape functions $\left(  \ref{b(r)CGM}\right)  $ and $\left(
\ref{b(r)CGMa}\right)  $ have been analyzed in Ref.\cite{BLLMM} and we know
that they will produce a \textit{Global Monopole}\cite{BV}. Nevertheless, an
interesting shape function can be obtained if one plugs the redshift function
$\left(  \ref{phi(r)b(r)}\right)  $ into Eq.$\left(  \ref{bprho}\right)  $.
Indeed, one finds%
\begin{equation}
b\left(  r\right)  =r\left(  1-{\frac{{\kappa{r}^{2}A}\left(  2\left(
8{\alpha}^{2}-12\alpha+1\right)  r_{0}^{2}-c\left(  r\right)  {\left(
r_{0}/r\right)  ^{\alpha}}\right)  }{\left(  3r+r_{0}\right)  ^{2}%
\alpha\left(  {\alpha}-1\right)  \left(  \alpha-2\right)  }}\right)
,\label{b(r)M}%
\end{equation}
where%
\begin{equation}
c\left(  r\right)  =\left(  3\,r+r_{0}\right)  ^{2}\alpha\left(  {\alpha
}-1\right)  -2\left(  r_{0}^{2}+3{r}r_{0}\right)  \alpha+2r_{0}^{2}{.}%
\end{equation}
When $r\rightarrow\infty$, we have the following asymptotic behavior for
$\alpha>2$%
\begin{equation}
b\left(  r\right)  \simeq r\left(  1-{\frac{2A\kappa r_{0}^{2}\left(
8{\alpha}^{2}-12\alpha+1\right)  }{9\alpha\,\left(  {\alpha}-1\right)  \left(
\alpha-2\right)  }}\right)  ,\label{Asb(r)}%
\end{equation}
which implies that, even in this case, we are in presence of a \textit{Global
Monopole}. Indeed, plugging the shape function $\left(  \ref{b(r)M}\right)  $
into the original metric $\left(  \ref{metric}\right)  $, we can write%
\begin{gather}
ds^{2}=-\,{\left(  \frac{4r}{3r+r_{0}}\right)  }^{2}dt^{2}+\frac{\left(
3r+r_{0}\right)  ^{2}\alpha\left(  {\alpha}-1\right)  \left(  \alpha-2\right)
dr^{2}}{{\kappa{r}^{2}A}\left(  2\left(  8{\alpha}^{2}-12\alpha+1\right)
r_{0}^{2}-c\left(  r\right)  {\left(  r_{0}/r\right)  ^{\alpha}}\right)
}+r^{2}\,(d\theta^{2}+\sin^{2}{\theta}\,d\varphi^{2})\,\nonumber\\
\underset{r\rightarrow\infty}{\simeq}-\frac{16}{9}\,dt^{2}+\frac
{9\alpha\,\left(  {\alpha}-1\right)  \left(  \alpha-2\right)  }{2A\kappa
r_{0}^{2}\left(  8{\alpha}^{2}-12\alpha+1\right)  }dr^{2}+r^{2}\,(d\theta
^{2}+\sin^{2}{\theta}\,d\varphi^{2})\nonumber\\
=\underset{r\rightarrow\infty}{\simeq}-\frac{16}{9}\,dt^{2}+d\tilde{r}%
^{2}+\Delta\tilde{r}^{2}\,(d\theta^{2}+\sin^{2}{\theta}\,d\varphi
^{2})\,,\label{dsGM}%
\end{gather}
where we have rescaled the radial coordinate and where we have defined%
\begin{equation}
\Delta=\frac{2A\kappa r_{0}^{2}\left(  8{\alpha}^{2}-12\alpha+1\right)
}{9\alpha\,\left(  {\alpha}-1\right)  \left(  \alpha-2\right)  }.
\end{equation}
If $\Delta>0$ we have an excess of the solid angle and this happens when%
\begin{equation}
\alpha>2;\qquad0<\alpha<\left(  3-\sqrt{7}\right)  /4;\qquad1<\alpha<\left(
3+\sqrt{7}\right)  /4.
\end{equation}
On the other hand when%
\begin{equation}
\alpha<0;\qquad\left(  3-\sqrt{7}\right)  /4<\alpha<1;\qquad\left(  3+\sqrt
{7}\right)  /4<\alpha<2
\end{equation}
we have a deficit of the solid angle, namely $\Delta<0$. The authors of
Ref.\cite{BLLMM} interpret the metric $\left(  \ref{b(r)CGM}\right)  $ as a
wormhole carrying a global monopole: the same interpretation can be applied to
the metric $\left(  \ref{dsGM}\right)  $. For completeness, we can compute the
SET components like the energy density%
\begin{equation}
\rho\left(  r\right)  =\frac{1}{\kappa r^{2}}\Bigg\{1-\frac{A\kappa\left(
\rho^{1}\left(  r\right)  \left(  r_{0}/{r}\right)  ^{\alpha}+\rho^{2}\left(
r\right)  \right)  }{9\alpha\,\left(  {\alpha}-1\right)  \left(
\alpha-2\right)  \left(  3r+r_{0}\right)  ^{2}}\Bigg\},
\end{equation}
where%
\begin{align}
\rho^{1}\left(  r\right)   &  =\left[  6{r}^{2}\left(  \left(  \alpha
-2\right)  \alpha\left(  6r+\alpha-4\right)  rr_{0}-9{r}^{2}\left(
\alpha+6r-5\right)  \left(  \alpha-1\right)  \alpha-\left(  6r+\alpha
-3\right)  \left(  \alpha-1\right)  \left(  \alpha-2\right)  r_{0}^{2}\right)
\right]  ,\\
\rho^{2}\left(  r\right)   &  =6\left(  2r-1\right)  \left(  8{\alpha}%
^{2}-12\alpha+1\right)  r_{0}^{2}.
\end{align}
The radial pressure is simply%
\begin{equation}
p_{r}\left(  r\right)  =-\frac{r}{\kappa r^{2}}\left(  1-{\frac{c\left(
r\right)  }{\left(  3r+r_{0}\right)  ^{2}\alpha\left(  {\alpha}^{2}%
-3\alpha+2\right)  }}\right)
\end{equation}
and finally the transverse pressure is%
\begin{equation}
p_{t}\left(  r\right)  =\frac{9\kappa\,A\left[  {p_{t}^{1}\left(  r\right)
}+{p_{t}^{2}\left(  r\right)  +p_{t}^{3}\left(  r\right)  +{4r_{0}^{4}\left(
8{\alpha}^{2}-12\alpha+1\right)  /3}}\right]  }{2\left(  {\alpha}-1\right)
\left(  \alpha-2\right)  \alpha\,\left(  3\,r+r_{0}\right)  ^{4}},
\end{equation}
where we have defined%
\begin{align}
p_{t}^{1}\left(  r\right)   &  =6\left(  \alpha-2\right)  \alpha r_{0}%
\,{r}\left(  \left(  {\frac{2\alpha-8}{9}}\right)  r_{0}^{2}+\left(
rr_{0}+{r}^{2}\right)  \left(  \alpha-1\right)  \right)  \left(  \frac{r_{0}%
}{r}\right)  ^{\alpha}{,}\\
p_{t}^{2}\left(  r\right)   &  =9\alpha{r}^{2}\left(  \left(  {\frac
{2\alpha-10}{9}}\right)  r_{0}^{2}+\left(  rr_{0}+{r}^{2}\right)  \left(
\alpha-2\right)  \right)  \left(  \alpha-1\right)  \,\left(  \frac{r_{0}}%
{r}\right)  ^{\alpha}{,}\\
p_{t}^{3}\left(  r\right)   &  =\left(  \alpha-2\right)  \left(
\alpha-1\right)  r_{0}^{2}\left(  \left(  \frac{2\alpha-6}{9}\right)
r_{0}^{2}+\alpha\,rr_{0}+\alpha\,{r}^{2}\right)  \left(  \frac{r_{0}}%
{r}\right)  ^{\alpha}%
\end{align}
On the \textit{throat}\ we find%
\begin{equation}
\rho\left(  r_{0}\right)  =\frac{1-A\kappa r_{0}^{2}}{\kappa r_{0}^{2}%
}>0\qquad\Longleftrightarrow\qquad1>A\kappa r_{0}^{2}{,}\qquad p_{r}\left(
r_{0}\right)  =-\frac{1}{\kappa r_{0}^{2}}<0,\qquad p_{t}(r_{0})={\frac{5A}%
{8}}>0.
\end{equation}
Once again, for the special case $\alpha=4$, one gets%
\begin{align}
\rho\left(  r\right)   &  ={\frac{\left(  -3A\kappa r_{0}^{2}+4\right)
{r}^{2}-A\kappa r_{0}^{4}}{4\kappa{r}^{4}}},\\
p_{r}\left(  r\right)   &  ={\frac{3\kappa{r}^{2}Ar_{0}^{2}-2A\kappa r_{0}%
^{3}r-A\kappa r_{0}^{4}-4{r}^{2}}{4\kappa r^{2}},}\\
p_{t}\left(  r\right)   &  ={\frac{\left(  9r+r_{0}\right)  r_{0}^{4}A}%
{4{r}^{4}\left(  3r+r_{0}\right)  }.}%
\end{align}

\section{The Return of the Traversable Wormhole and the Disappearance of the
Global Monopole}

\label{p4}The throat condition constrains the shape function to have a fixed
point, but not its value. Thanks to this freedom, we can reexamine the shape
functions obtained previously and we show that for appropriate choices of the
parameters, a TW without a monopole can appear. To this purpose, let us begin
from Eq.$\left(  \ref{b(r)CGM}\right)  $ and by observing that nothing forbids
the following identification%
\begin{equation}
1-\frac{\kappa Ar_{0}^{\alpha}}{\alpha-2}\frac{1}{r_{0}^{\alpha-2}}%
=0\qquad\Longleftrightarrow\qquad r_{0}=\sqrt{\frac{\alpha-2}{\kappa A}%
},\label{NoGM}%
\end{equation}
one obtains the following relationship
\begin{equation}
b\left(  r\right)  =\frac{r_{0}^{\alpha-2}}{r^{\alpha-3}},\qquad\alpha
\neq2.\label{b(r)alpha}%
\end{equation}
Note that the assumption $\left(  \ref{NoGM}\right)  $ is not possible when
$\alpha=2$. Since the shape function $\left(  \ref{b(r)alpha}\right)  $ has
been obtained imposing ZTF, the computation of the SET is immediate and we get%
\begin{equation}
\rho\left(  r\right)  =-\left(  \alpha-3\right)  \frac{r_{0}^{\alpha-2}%
}{\kappa r^{\alpha}},\qquad p_{r}\left(  r\right)  =-\frac{r_{0}^{\alpha-2}%
}{\kappa r^{\alpha}},\qquad p_{t}(r)=\left(  \alpha-2\right)  \frac
{r_{0}^{\alpha-2}}{2\kappa r^{\alpha}}.
\end{equation}
For the particular case of $\alpha=4$, we recover the EB wormhole and a direct
comparison with the Casimir SET allows to fix also the constant $A$. Indeed,
one finds%
\begin{equation}
\rho\left(  r\right)  =-\frac{r_{0}^{2}}{\kappa r^{4}}=-\frac{\hbar c\pi^{2}%
}{720r^{4}}\qquad\Longrightarrow\qquad r_{0}^{2}=\frac{\pi^{3}l_{P}^{2}}%
{90}\qquad\Longrightarrow\qquad A=\frac{45\hbar c}{2l_{P}^{4}\pi^{4}}.
\end{equation}
For such an identification, the final SET will be equal to the SET $\left(
\ref{TEB1}\right)  $ and $r_{0}=r_{1}$. However, it is immediate to recognize
that the energy density $A$ is so huge that cannot have a physical meaning.
Moreover if we also impose%
\begin{equation}
A=-\frac{4\hbar c\pi^{2}}{720a^{4}}\qquad\Longrightarrow\qquad a=0.7l_{P},
\end{equation}
where $a$ is the fixed plates distance. This further constraint is even worst
than what has been found in Ref.\cite{MTY}. Nevertheless if we constrain only
the relationship $\left(  \ref{NoGM}\right)  $,%
\begin{equation}
r_{0}=\sqrt{\frac{\alpha-2}{\kappa A}}=\frac{a^{2}}{l_{P}}\sqrt{\frac
{90\left(  \alpha-2\right)  }{4\pi^{3}}}.
\end{equation}
and we consider a standard Casimir separation with $a\simeq1\mu m$ and
$\alpha=4$, we find an opposite huge result%
\begin{equation}
r_{0}\simeq\allowbreak3.\,\allowbreak727\,3\times10^{22}%
\operatorname{m}%
.
\end{equation}
We need to reach a plate separation of the order of the $fm,$to have a throat
of the order of $10^{5}m$. One could be tempted to use the arbitrariness of
$\alpha$. However $\alpha>3$, otherwise the energy density becomes positive.
The other case where the TW returns and Global Monopole disappears is
represented by the shape function $\left(  \ref{b(r)M}\right)  $, whose
asymptotic behavior is represented by Eq.$\left(  \ref{Asb(r)}\right)  $. With
the assumption%
\begin{equation}
A={\frac{9\alpha\left(  {\alpha-}2\right)  \left(  {\alpha-1}\right)
}{2\kappa r_{0}^{2}\left(  8{\alpha}^{2}-12\alpha+1\right)  },}%
\end{equation}
one finds the TW shape function $\forall\alpha>2$%
\begin{equation}
b\left(  r\right)  =r\left(  1-{\frac{9{{r}^{2}}\left(  2\left(  8{\alpha}%
^{2}-12\alpha+1\right)  r_{0}^{2}-c\left(  r\right)  {\left(  r_{0}/r\right)
^{\alpha}}\right)  }{2r_{0}^{2}\left(  8{\alpha}^{2}-12\alpha+1\right)
\left(  3r+r_{0}\right)  ^{2}}}\right)  .
\end{equation}
In particular for $\alpha=4$, one finds
\begin{equation}
b\left(  r\right)  =\frac{r_{0}}{3\left(  3r+r_{0}\right)  ^{2}{{{r}}}}\left[
\allowbreak\left(  2r+r_{0}\right)  \left(  3r+r_{0}\right)  ^{2}\right]
=\frac{2r_{0}}{3}+\frac{r_{0}^{2}}{3r}%
\end{equation}
in agreement with the shape function described in $\left(  \ref{phi(r)b(r)}%
\right)  $. The SET coincides with the one described by $\left(
\ref{Tmn2}\right)  $. In particular one finds%
\begin{equation}
\rho\left(  r\right)  =-{\frac{{\pi}^{2}\hbar c}{720{r}^{4}}},\qquad
p_{r}\left(  r\right)  =-{\frac{3{\pi}^{2}\hbar c}{720{r}^{4}}}\qquad\qquad
p_{t}(r)=\frac{{\pi}^{2}\hbar c}{720{r}^{4}}\left(  {\frac{9r+r_{0}}{3r+r_{0}%
}}\right)  {,}%
\end{equation}
where we have fine-tuned the parameters in such a way to produce the correct
values of the Casimir energy density and pressure. As one can see, the
disagreement with the transverse pressure persists. If we put the physical
numbers we have used for the $\left(  \ref{b(r)alpha}\right)  $ case, we find
no substantial difference with that case.

\section{Conclusions}

\label{p5}In this paper, we have extended the study began by Morris, Thorne
and Yurtsever in Ref.\cite{MTY} and subsequently explored by
Visser\cite{Visser} on the Casimir effect as a possible source for a TW. By
imposing an EoS, we have discovered a solution depending on the ratio between
the throat radius and $r_{1}$ which is directly connected with the Planck
length. As shown in Eq.$\left(  \ref{or1r0}\right)  $ if $\omega>\omega_{0}$,
we have a black hole, while if $\omega=\omega_{0}$, we have a TW. Note also
that for $\omega<\omega_{0}$, we have a singularity has shown in app.\ref{A2}.
For the TW case, one finds that for the special value $\omega=3$, the size of
the TW is Planckian and therefore it is traversable in principle but not in
practice. The TW $\left(  \ref{dSC2}\right)  $ reproduces the original Casimir
energy density $\left(  \ref{rhoC}\right)  $ and pressure $\left(
\ref{P(a)}\right)  $, but the transverse pressure is in disagreement with the
one associated with the TW $\left(  \ref{dSC2}\right)  $. This is a
consequence of having assumed a variable separation between the plates instead
of a constant separation like the original Casimir effect. Therefore the
correspondence between the original Casimir SET and the one computed by means
of the line element $\left(  \ref{dSC2}\right)  $ is not one to one.
Nevertheless, it is remarkable that a solution connecting the Casimir energy
and a TW has been found because, as far as we know, in the literature nothing
is said about this point. It is also remarkable that the same EoS with
$\omega=1$ is able to reproduce the EB wormhole, but even in this case the
correspondence with the Casimir SET is not one to one. For this reason we have
explored the possibility of supporting the TW $\left(  \ref{dSC2}\right)  $
with other forms of constraint. An interesting condition comes from the QWEC
which, in general, produces a global monopole. However, even in this case it
is always possible to use the arbitrariness of the throat and eliminate the
global monopole in favor of an appearing TW. Unfortunately, even if the QWEC
corroborates the results obtained in section \ref{p2}, when one puts numbers
inside the parameters, the expectations are far to be momentarily interesting.
A comment about the EoS with $\omega=3$ is in order. Since it is the NEC that
must be violated, many proposals to keep $\rho\left(  r\right)  >0$ have been
done\cite{phantomWH1,phantomWH2,phantomWH3}. However, even if the results are
encouraging, we have to say that there is no knowledge on how to build
\textquotedblleft\textit{phantom energy}\textquotedblright\ in practice. I
recall that phantom energy obeys the following relationship%
\begin{equation}
p_{r}\left(  r\right)  =\omega\rho\left(  r\right)  ,\qquad\Longrightarrow
\qquad p_{r}\left(  r\right)  +\rho\left(  r\right)  <0\qquad
\Longleftrightarrow\qquad\left(  1+\omega\right)  \rho\left(  r\right)  <0,
\end{equation}
which implies%
\begin{equation}
\rho\left(  r\right)  >0,\qquad\Longrightarrow\qquad\omega<-1.
\end{equation}
On this ground the Casimir effect offers a viable interesting realizable model
of \textit{exotic matter} which, however, can contribute only at the Planck
scale. Always on the side of phantom energy I proposed the idea of
\textit{Self-Sustained Traversable Wormholes}, namely TW sustained by their
own quantum fluctuations\cite{RG,RG1,RGFSNL,RGFSNL1,RGFSNL2}. Even in this
case, because the quantum fluctuation carried by the graviton behaves like the
ordinary Casimir effect, we found that no need for phantom contribution is
necessary. On this context, in a next paper we will explore how behaves a
system formed by the \textit{Casimir TW}, here analyzed, and the corresponding
self-sustained TW version.

\section{Acknowledgments}

I would like to thank Francisco S.N. Lobo for useful discussions and
suggestions for the Traversable Wormhole part and I would like to thank Enrico
Calloni and Luigi Rosa for the Casimir effect.

\appendix{}

\section{Properties of the Casimir wormhole}

\label{A1}In section \ref{p2}, we have introduced the shape function $\left(
\ref{dSC3}\right)  $ obtained by the Casimir energy. Here we want to discuss
some of its properties, even if the wormhole has Planckian size. The first
quantity we are going to analyze is the proper radial distance which is
related to the shape function by%
\begin{align}
l\left(  r\right)   &  =\pm\int_{r_{0}}^{r}\frac{dr^{\prime}}{\sqrt
{1-\frac{b\left(  r^{\prime}\right)  }{r^{\prime}}}}=\pm\int_{r_{0}}^{r}%
\frac{dr^{\prime}}{\sqrt{1-\frac{2r_{0}}{3r^{\prime}}-\frac{r_{0}^{2}%
}{3r^{\prime2}}}}\nonumber\\
=\pm &  \frac{\sqrt{3}\sqrt{\left(  3r+r_{0}\right)  \left(  r-r_{0}\right)
}}{3}+\frac{r_{0}}{3}\ln\left(  \frac{3r-r_{0}+\sqrt{3}\sqrt{\left(
3\,r+r_{0}\right)  \left(  r-r_{0}\right)  }}{2r_{0}}\right)  .\label{l(r)}%
\end{align}
We find%
\begin{equation}
l\left(  r\right)  \simeq%
\begin{array}
[c]{ll}%
\pm\left(  r-r_{0}/3\left(  1-\ln\left(  3r/r_{0}\right)  \right)  +O\left(
\frac{1}{r}\right)  \right)   & r\rightarrow\infty\\
\pm\sqrt{3r_{0}\left(  r-r_{0}\right)  }+O\left(  \left(  r-r_{0}\right)
^{\frac{3}{2}}\right)   & r\rightarrow r_{0}%
\end{array}
,
\end{equation}
where the\textquotedblleft$\pm$\textquotedblright\ depends on the wormhole
side we are. The proper radial distance is an essential tool to estimate the
possible time trip in going from one station located in the lower universe,
say at $l=-l_{1}$, and ending up in the upper universe station, say at
$l=l_{2}$. Following Ref.\cite{MT}, we shall locate $l_{1}$ and $l_{2}$ at a
value of the radius such that $l_{1}\simeq l_{2}\simeq10^{4}r_{0}$ that it
means $1-b\left(  r\right)  /r\simeq1$. Assume that the traveller has a radial
velocity $v\left(  r\right)  $, as measured by a static observer positioned at
$r$. One may relate the proper distance travelled $dl$, radius travelled $dr$,
coordinate time lapse $dt$, and proper time lapse as measured by the observer
$d\tau$, by the following relationships%
\begin{equation}
v=e^{-\phi\left(  r\right)  }\frac{dl}{dt}=e^{-\phi\left(  r\right)  }\left(
1-\frac{b\left(  r\right)  }{r}\right)  ^{-\frac{1}{2}}\frac{dr}{dt}\label{vt}%
\end{equation}
and%
\begin{equation}
v\gamma=\frac{dl}{d\tau}=\mp\left(  1-\frac{b\left(  r\right)  }{r}\right)
^{-\frac{1}{2}}\frac{dr}{d\tau};\qquad\gamma=\left(  1-\frac{v^{2}\left(
r\right)  }{c^{2}}\right)  ^{-\frac{1}{2}}\label{vtau}%
\end{equation}
respectively. If the traveler journeys with constant speed $v$, then the total
time is given by%
\begin{equation}
\Delta t=\int_{r_{0}}^{r}\frac{e^{-\phi\left(  r^{\prime}\right)  }dr^{\prime
}}{v\sqrt{1-\frac{b\left(  r^{\prime}\right)  }{r^{\prime}}}}=\frac{1}%
{4v}\sqrt{9{r}^{2}-6r_{0}r-3r_{0}^{2}}+\frac{r_{0}}{2v}\ln\left(  \frac
{3\,r}{2r_{0}}+\frac{\sqrt{9{r}^{2}-6rr_{0}-3r_{0}^{2}}}{2r_{0}}-\frac{1}%
{2}\right)  ,
\end{equation}
while the proper total time is%
\begin{equation}
\Delta\tau=\int_{r_{0}}^{r}\frac{dr^{\prime}}{v\sqrt{1-\frac{b\left(
r^{\prime}\right)  }{r^{\prime}}}}=\frac{1}{3v}\sqrt{9{r}^{2}-6rr_{0}%
-3r_{0}^{2}}+\frac{r_{0}}{3v}\ln\left(  \frac{3\,r}{2r_{0}}+\frac{\sqrt
{9{r}^{2}-6rr_{0}-3r_{0}^{2}}}{2r_{0}}-\frac{1}{2}\right)  .
\end{equation}
As we can see $\Delta\tau\simeq\Delta t$. On the same ground, we can compute
the embedded surface, which is defined by%
\begin{equation}
z\left(  r\right)  =\pm\int_{r_{0}}^{r}\frac{dr^{\prime}}{\sqrt{\frac{b\left(
r^{\prime}\right)  }{r^{\prime}}-1}}%
\end{equation}
and, in the present case, we find%
\begin{align}
z\left(  r\right)   &  =\pm\sqrt{\frac{r_{0}}{3}}\int_{r_{0}}^{r}\frac
{\sqrt{2r^{\prime}+r_{0}}dr^{\prime}}{\sqrt{\left(  r^{\prime}-r_{0}\right)
\left(  r^{\prime}+r_{0}/3\right)  }}\\
&  =\frac{2r_{0}}{9}\left(  F\left(  \sqrt{3}\sqrt{{\frac{r-r_{0}}{r_{0}%
+3\,r}}},\frac{1}{3}\right)  +8\Pi{\left(  \sqrt{3}\sqrt{{\frac{r-r_{0}}%
{r_{0}+3\,r}}},1,\frac{1}{3}\right)  }\right)  ,
\end{align}
\begin{figure}[tbh]
\centering
\includegraphics[width=2.8in]{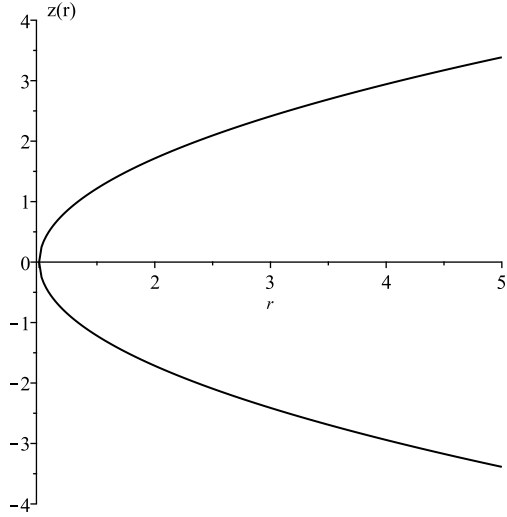}\caption{Representation of the
embedding diagram}%
\end{figure}

where $F\left(  \varphi,k\right)  $ is the elliptic integral of the first kind
and $\Pi\left(  \varphi,n,k\right)  $ is the elliptic integral of the third
kind. To further investigate the properties of the shape function $\left(
\ref{dSC3}\right)  $, we consider the computation of the total gravitational
energy for a wormhole\cite{NZCP}, defined as%
\begin{equation}
E_{G}\left(  r\right)  =\int_{r_{0}}^{r}\left[  1-\sqrt{\frac{1}{1-b\left(
r^{\prime}\right)  /r^{\prime}}}\right]  \rho\left(  r^{\prime}\right)
dr^{\prime}r^{\prime2}+\frac{r_{0}}{2G}=M-M_{\pm}^{P},
\end{equation}
where $M$ is the total mass $M$ and $M^{P}$ is the proper mass, respectively.
Even in this case, the \textquotedblleft$\pm$\textquotedblright\ depends one
the wormhole side we are. In particular%
\begin{equation}
M=\int_{r_{0}}^{r}4\pi\rho\left(  r^{\prime}\right)  r^{\prime2}dr^{\prime
}+\frac{r_{0}}{2G}=\frac{1}{2G}\left(  \frac{2r_{0}}{3}+\frac{r_{0}^{2}}%
{3r}-r_{0}\right)  +\frac{r_{0}}{2G}=\frac{1}{2G}\left(  \frac{2r_{0}}%
{3}+\frac{r_{0}^{2}}{3r}\right)  \underset{r\rightarrow\infty}{\simeq}%
\frac{r_{0}}{3G}\label{M}%
\end{equation}
and%
\begin{align}
M_{\pm}^{P} &  =\pm\int_{r_{0}}^{r}\frac{4\pi\rho\left(  r^{\prime}\right)
r^{\prime2}}{\sqrt{1-b\left(  r^{\prime}\right)  /r^{\prime}}}dr^{\prime}%
=\mp\frac{\sqrt{3}c^{4}r_{0}^{2}}{6G}\int_{r_{0}}^{r}\frac{dr^{\prime}%
}{r^{\prime}\sqrt{3r^{\prime2}-6r_{0}r^{\prime}-r_{0}^{2}}}\label{Mp}\\
&  =\mp{\frac{\sqrt{3}c^{4}r_{0}}{12G}\left(  \pi-2\,\arctan\left(
{\frac{r_{0}+r}{\sqrt{3{r}^{2}-2r_{0}r-r_{0}^{2}}}}\right)  \right)
}\underset{r\rightarrow\infty}{\simeq}\mp{\frac{\sqrt{3}{\pi}c^{4}r_{0}}{18G}%
}.
\end{align}
An important traversability condition is that the acceleration felt by the
traveller should not exceed Earth's gravity $g_{\oplus}\simeq980$ $cm/s^{2}$.
In an orthonormal basis of the traveller's proper reference frame, we can find%
\begin{equation}
\left\vert \mathbf{a}\right\vert =\left\vert \sqrt{1-\frac{b\left(  r\right)
}{r}}e^{-\phi\left(  r\right)  }\left(  \gamma e^{\phi\left(  r\right)
}\right)  ^{\prime}\right\vert \leq\frac{g_{\oplus}}{c^{2}}.
\end{equation}
If we assume a constant speed and $\gamma\simeq1$, then we can write%
\begin{equation}
\left\vert \mathbf{a}\right\vert =\left\vert \sqrt{1-\frac{2r_{0}}{3r}%
-\frac{r_{0}^{2}}{3r}}\frac{r_{0}}{r\left(  3r+r_{0}\right)  }\right\vert
\leq\frac{g_{\oplus}}{c^{2}}.
\end{equation}
We can see that in proximity of the throat, the traveller has no acceleration.
Always following Ref.\cite{MT}, we can estimate the tidal forces by imposing
an upper bound represented by $g_{\oplus}$. The radial tidal constraint
\begin{equation}
\left\vert \left(  1-\frac{b\left(  r\right)  }{r}\right)  \left[
\phi^{\prime\prime}\left(  r\right)  +\left(  \phi^{\prime}\left(  r\right)
\right)  ^{2}-\frac{b^{\prime}\left(  r\right)  r-b\left(  r\right)
}{2r\left(  r-b\left(  r\right)  \right)  }\phi^{\prime}\left(  r\right)
\right]  c^{2}\right\vert \left\vert \eta^{\hat{1}^{\prime}}\right\vert \leq
g_{\oplus},\label{RTC}%
\end{equation}
constrains the redshift function, and the lateral tidal constraint%
\begin{equation}
\left\vert \frac{\gamma^{2}c^{2}}{2r^{2}}\left[  \frac{v^{2}\left(  r\right)
}{c^{2}}\left(  b^{\prime}\left(  r\right)  -\frac{b\left(  r\right)  }%
{r}\right)  +2r\left(  r-b\left(  r\right)  \right)  \phi^{\prime}\left(
r\right)  \right]  \right\vert \left\vert \eta^{\hat{2}^{\prime}}\right\vert
\leq g_{\oplus},\label{LTC}%
\end{equation}
constrains the velocity with which observers traverse the wormhole.
$\eta^{\hat{1}^{\prime}}$ and $\eta^{\hat{2}^{\prime}}$ represent the size of
the traveller. In Ref.\cite{MT}, they are fixed approximately equal, at the
symbolic value of $2$ $m$. Close to the throat, the radial tidal constraint
$\left(  \ref{RTC}\right)  $ becomes\footnote{Note that if we put the
Planckian value in $r_{0}$ obtained in $\left(  \ref{phi(r)b(r)}\right)  $,
then one finds $\left\vert \eta^{\hat{1}^{\prime}}\right\vert \leq
1.\,\allowbreak057\,3\times10^{-86}%
\operatorname{m}%
$. This means that with a Planckian wormhole nothing can traverse it.}%
\begin{equation}
\left\vert \left[  \frac{b\left(  r\right)  -b^{\prime}\left(  r\right)
r}{2r^{2}}\phi^{\prime}\left(  r\right)  \right]  \right\vert \leq
\frac{g_{\oplus}}{c^{2}\left\vert \eta^{\hat{1}^{\prime}}\right\vert }%
\simeq\left(  10^{8}m\right)  ^{-2}\qquad\Longrightarrow\qquad10^{8}m\lesssim
r_{0}.\label{RTCt}%
\end{equation}
For the lateral tidal constraint, we find%
\begin{equation}
\frac{v^{2}r_{0}}{3r^{4}}\left(  r+r_{0}\right)  \left\vert \eta^{\hat
{2}^{\prime}}\right\vert \lesssim g_{\oplus}\qquad\Longrightarrow\qquad
v\lesssim r_{0}\sqrt{\frac{3g_{\oplus}}{4}}\qquad\Longrightarrow\qquad
v\lesssim2.7r_{0}\text{ }m/s.\label{LTCt}%
\end{equation}
If the observer has a vanishing $v$, then the tidal forces are null. We can
use these last estimates to complete the evaluation of the crossing time which
approximately is%
\begin{equation}
\Delta t\simeq2\times10^{4}\frac{3r_{0}}{4v}\simeq5\times10^{3}s,
\end{equation}
which is in agreement with the estimates found in Ref.\cite{MT}.The last
property we are going to discuss is the \textquotedblleft total
amount\textquotedblright\ of ANEC violating matter in the spacetime\cite{VKD}
which is described by Eq. $\left(  \ref{IV}\right)  $. For the metric $\left(
\ref{dSC3}\right)  $, one obtains
\begin{equation}
I_{V}=-\frac{1}{\kappa}\int_{r_{0}}^{\infty}\frac{4r_{0}^{2}}{3r^{2}}%
dr=-\frac{4r_{0}}{3\kappa}.
\end{equation}
Differently from what we have computed in Eq.$\left(  \ref{IVmw0}\right)  $,
this time the result is finite everywhere. Therefore we can conclude that, in
proximity of the throat the ANEC can be arbitrarily small.

\section{A Particular case: transforming a singularity into a TW}

\label{A2}As a sub-case of the solution $\left(  \ref{b(r)0}\right)  $, we
consider the following assumption on the throat radius%
\begin{equation}
r_{0}=r_{1}%
\end{equation}
representing the EB wormhole shape function. Plugging $b\left(  r\right)
=r_{0}^{2}/r$ and Eq.$\left(  \ref{P(a)}\right)  $ into Eq.$\left(
\ref{pr0}\right)  $, we find the following form for the redshift function%
\begin{equation}
\phi\left(  r\right)  =-\frac{1}{2}\ln\left(  1-\frac{r_{0}^{2}}{r^{2}%
}\right)  +\phi\left(  r_{0}\right)  ,
\end{equation}
with the usual replacement of $a$ with $r$. We find that in $r=r_{0}$ we have
neither a black hole nor a traversable wormhole but we have a singularity
described by the following line element%
\begin{equation}
ds^{2}=-\,\frac{dt^{2}}{{1-\frac{r_{0}^{2}}{r^{2}}}}+\frac{dr^{2}}%
{1-\frac{r_{0}^{2}}{r^{2}}}+r^{2}\,(d\theta^{2}+\sin^{2}{\theta}\,d\varphi
^{2})\,.\label{Sing}%
\end{equation}
The metric $\left(  \ref{Sing}\right)  $ has been obtained also in
Ref.\cite{Lobo} with $\alpha=-1$ and $\omega=3$, out of the phantom region.
Because of $\left(  \ref{Phi(r)2}\right)  $, $p_{t}(r)$ is divergent when
$r\rightarrow r_{0}$; indeed from the line element $\left(  \ref{Sing}\right)
$, we find\cite{Kim}%
\begin{equation}
p_{t}(r)={\frac{\left(  3{r}^{2}-r_{0}^{2}\right)  r_{0}^{2}}{{r}^{4}\left(
{r}^{2}-r_{0}^{2}\right)  }.}%
\end{equation}
The presence of a singularity for the metric $\left(  \ref{Sing}\right)  $ is
also confirmed by the calculation of the Kretschmann scalar%
\begin{equation}
R_{\alpha\beta\gamma\delta}R^{\alpha\beta\gamma\delta}=8{\frac{r_{0}%
^{4}\left(  7{r}^{4}-8{r}^{2}r_{0}^{2}+3r_{0}^{4}\right)  }{\left(
r-r_{0}\right)  ^{2}\left(  r+r_{0}\right)  ^{2}{r}^{8}}.}%
\end{equation}
Nevertheless, instead of discarding the metric $\left(  \ref{Sing}\right)  $,
we can use the strategy adopted in Refs.\cite{DS,Visser} and we slightly
modify the line element $\left(  \ref{Sing}\right)  $ in the following way%
\begin{equation}
ds^{2}=-\,\frac{dt^{2}}{{1-\frac{{\lambda}r_{0}^{2}}{r^{2}}}}+\frac{dr^{2}%
}{1-\frac{r_{0}^{2}}{r^{2}}}+r^{2}\,(d\theta^{2}+\sin^{2}{\theta}%
\,d\varphi^{2})\,,\qquad{\lambda}\in\left[  0,1\right]  .\label{NoSing}%
\end{equation}
Of course, we could have adopted the following distortion of the metric
$\left(  \ref{Sing}\right)  $%
\begin{equation}
ds^{2}=-\,\frac{dt^{2}}{{1-\frac{r_{0}^{2}}{r^{2}}+\lambda}}+\frac{dr^{2}%
}{1-\frac{r_{0}^{2}}{r^{2}}}+r^{2}\,(d\theta^{2}+\sin^{2}{\theta}%
\,d\varphi^{2}),\label{NoSing1}%
\end{equation}
like in Refs.\cite{DS,Visser}. However, the line element $\left(
\ref{NoSing}\right)  $ has the advantage of interpolating between the
singularity, when ${\lambda}=1$, and the EB wormhole when ${\lambda}=0$, while
the line element $\left(  \ref{NoSing1}\right)  $ has not this property. This
simple modification produces the following effects on the SET%
\begin{equation}
T_{\mu\nu}=diag\left(  \rho\left(  r\right)  ,p_{r}^{{\lambda}}(r),p_{t}%
^{{\lambda}}(r),p_{t}^{{\lambda}}(r)\right)  ,
\end{equation}
where%
\begin{align}
\rho\left(  r\right)   &  =-\frac{r_{0}^{2}}{\kappa r^{4}},\\
p_{r}^{{\lambda}}(r) &  =-\frac{r_{0}^{2}}{\kappa r^{4}}\frac{{r}^{2}\left(
2{\lambda+1}\right)  -3{\lambda}r_{0}^{2}}{{r}^{2}-{\lambda}r_{0}^{2}},\\
p_{t}^{{\lambda}}(r) &  =\frac{r_{0}^{2}}{\kappa r^{4}}{\frac{\left(
2{\lambda}+1\right)  {r}^{4}+r_{0}^{2}{\lambda}\left(  {\lambda}-5\right)
{r}^{2}+\,r_{0}^{4}{\lambda}^{2}}{\left(  {r}^{2}-{\lambda}r_{0}^{2}\right)
^{2}}.}%
\end{align}
In proximity of the throat, we find%
\begin{align}
&  \rho\left(  r\right)  \underset{r\rightarrow r_{0}}{\rightarrow}-\frac
{1}{\kappa r_{0}^{2}},\\
&  p_{r}^{{\lambda}}(r)\underset{r\rightarrow r_{0}}{\rightarrow}-\frac
{1}{\kappa r_{0}^{2}},\\
&  p_{t}^{{\lambda}}(r)\underset{r\rightarrow r_{0}}{\rightarrow}%
{\frac{2{\lambda}^{2}-3{\lambda+1}}{\kappa r_{0}^{2}\left(  1-{\lambda
}\right)  ^{2}}}%
\end{align}
and for small ${\lambda}$, we find%
\begin{align}
p_{r}^{{\lambda}}(r) &  \simeq-{\frac{r_{0}^{2}}{\kappa{r}^{4}}}%
-{\frac{2{\lambda}r_{0}^{2}}{\kappa{r}^{4}}}+{\frac{2r_{0}^{4}{\lambda}%
}{\kappa{r}^{6}}+O}\left(  {\lambda}^{2}\right)  ,\\
p_{t}^{{\lambda}}(r) &  \simeq{\frac{r_{0}^{2}}{\kappa{r}^{4}}}+{\frac
{2{\lambda}r_{0}^{2}}{\kappa{r}^{4}}}-{\frac{3r_{0}^{4}{\lambda}}{\kappa
{r}^{6}}+O}\left(  {\lambda}^{2}\right)  .
\end{align}
If we assume that the relationship $r_{0}=r_{1}$ holds, one finds%
\begin{equation}
\frac{r_{0}^{2}}{\kappa}=\frac{\hbar c\pi^{2}}{720},
\end{equation}
i.e. the Casimir coefficients. Note that%
\begin{equation}
\lim_{{{\lambda\rightarrow0}}}\lim_{r\rightarrow r_{0}}T_{\mu\nu}%
=\lim_{r\rightarrow r_{0}}\lim_{{{\lambda\rightarrow0}}}T_{\mu\nu},
\end{equation}
while%
\begin{equation}
\lim_{{{\lambda\rightarrow1}}}\lim_{r\rightarrow r_{0}}T_{\mu\nu}\neq
\lim_{r\rightarrow r_{0}}\lim_{{{\lambda\rightarrow1}}}T_{\mu\nu},
\end{equation}
Indeed%
\begin{equation}
\frac{1}{\kappa r_{0}^{2}}\lim_{{{\lambda\rightarrow1}}}diag\left(
-1,-1{,\frac{2{\lambda-1}}{{\lambda-1}},\frac{2{\lambda-1}}{{\lambda-1}}%
}\right)  \neq\frac{1}{\kappa r_{0}^{2}}\lim_{r\rightarrow r_{0}}diag\left(
-1,-3{,1-{\frac{2r_{0}^{2}{r}^{2}}{\left(  {r}^{2}-r_{0}^{2}\right)  ^{2}}}%
,}1-{\frac{2r_{0}^{2}{r}^{2}}{\left(  {r}^{2}-r_{0}^{2}\right)  ^{2}}}\right)
\end{equation}
and the divergence persists in $p_{t}(r)$ but with a different behavior. The
same problem appears in Ref.\cite{DS}, where the authors examine a line
element of the form%
\begin{equation}
ds^{2}=-\left(  {1-\frac{r_{0}}{r}+\lambda}\right)  \,dt^{2}+\frac{dr^{2}%
}{1-\frac{r_{0}}{r}}+r^{2}\,(d\theta^{2}+\sin^{2}{\theta}\,d\varphi^{2})\,,
\end{equation}
leading to the following SET%
\begin{equation}
T_{\mu\nu}=\frac{r_{0}}{\kappa{r}^{2}}diag\left(  0,-{\frac{{\lambda}%
}{{\lambda}r+r-r_{0}},{\frac{\left(  2{\lambda}r+2r-r_{0}\right)  {\lambda}%
}{4\left(  {\lambda}r+r-r_{0}\right)  ^{2}}},}\frac{\left(  2{\lambda
}r+2r-r_{0}\right)  {\lambda}}{4\left(  {\lambda}r+r-r_{0}\right)  ^{2}%
}\right)  .
\end{equation}
It is easy to check that even in this case, a non-commutative behavior
appears. Indeed%
\begin{equation}
\lim_{{{\lambda\rightarrow0}}}T_{\mu\nu}=\frac{1}{\kappa{r}_{0}}diag\left(
0,-{\frac{{1}}{r_{0}},{\frac{{\lambda}+1}{2{\lambda}r_{0}}},}\frac{{\lambda
}+1}{2{\lambda}r_{0}}\right)  \neq\lim_{r\rightarrow r_{0}}\lim_{{{\lambda
\rightarrow0}}}T_{\mu\nu}=\frac{1}{\kappa{r}_{0}}diag\left(  0,0{,0,}0\right)
.
\end{equation}

\end{document}